\documentclass[pdflatex,sn-nature, iicol]{sn-jnl}

\usepackage{graphicx}%
\usepackage{multirow}%
\usepackage{amsmath,amssymb,amsfonts}%
\usepackage{amsthm}%
\usepackage{mathrsfs}%
\usepackage[title]{appendix}%
\usepackage{xcolor}%
\usepackage{textcomp}%
\usepackage{manyfoot}%
\usepackage{booktabs}%
\usepackage{algorithm}%
\usepackage{algorithmicx}%
\usepackage{algpseudocode}%
\usepackage{listings}%

\usepackage{siunitx}
\usepackage{layout}
\usepackage[]{geometry}
\usepackage{float}

\theoremstyle{thmstyleone}%
\theoremstyle{thmstyletwo}%

\theoremstyle{thmstylethree}%

\begin{document}

\title[Article Title]{Triggered telecom C-band single-photon source with high brightness, high indistinguishability and sub-GHz spectral linewidth}


\author*[1]{\fnm{Raphael} \sur{Joos}}\email{r.joos@ihfg.uni-stuttgart.de}
\author[1]{\fnm{Stephanie} \sur{Bauer}}
\author[1]{\fnm{Christian} \sur{Rupp}}
\author[1]{\fnm{Sascha} \sur{Kolatschek}}
\author[1]{\fnm{Wolfgang} \sur{Fischer}}
\author[1]{\fnm{Cornelius} \sur{Nawrath}}
\author[1]{\fnm{Ponraj} \sur{Vijayan}}
\author[1]{\fnm{Robert} \sur{Sittig}}


\author[1]{\fnm{Michael} \sur{Jetter}}
\author[1]{\fnm{Simone L.} \sur{Portalupi}}
\author[1]{\fnm{Peter} \sur{Michler}}

\affil[1]{\orgdiv{Institut f\"ur Halbleiteroptik und Funktionelle Grenzfl\"achen (IHFG), Center for Integrated Quantum Science and Technology (IQ$^{ST}$) and SCoPE}, \orgname{University of Stuttgart}, \orgaddress{\street{Allmandring 3}, \city{Stuttgart}, \postcode{70569}, \country{Germany}}}




\abstract{Long-range, terrestrial quantum networks will require high brightness single-photon sources emitting in the telecom C-band for maximum transmission rate. Many applications additionally demand triggered operation with high indistinguishability and narrow spectral linewidth. This would enable the efficient implementation of photonic gate operations and photon storage in quantum memories, as for instance required for a quantum repeater. Especially, semiconductor quantum dots (QDs) have shown these properties in the near-infrared regime. However, the simultaneous demonstration of all these properties in the telecom C-band has been elusive. Here, we present a coherently (incoherently) optically-pumped narrow-band (\SI{0.8}{\giga\hertz}) triggered single-photon source in the telecom C-band. The source shows simultaneously high single-photon purity with $g^{(2)}(0) = \SI{0.026}{}$ ($g^{(2)}(0) = \SI{0.014}{}$), high two-photon interference visibility of \SI{0.508}{} (\SI{0.664}{}) and high application-ready rates of \SI{0.75}{\mega\hertz} (\SI{1.45}{\mega\hertz}) of polarized photons. The source is based on a QD coupled to a circular Bragg grating cavity combined with spectral filtering. Coherent (incoherent) operation is performed via the novel SUPER scheme (phonon-assisted excitation).}






\maketitle

A key building block of many prospective quantum technologies such as quantum key distribution (QKD) \cite{Bozzio2022,Xu2020} or quantum metrology \cite{Chu2017,Georgieva2021} are highly efficient sources of single photons. Semiconductor quantum dot (QD) based single-photon sources are commonly viewed as a promising platform for those applications especially due to their exceptional brightness \cite{Tomm2021, Wang2019} in combination with a low multi-photon probability. However, as many applications deal with the distribution of information on a metropolitan or even larger scale, minimizing the losses in a quantum link becomes essential. For this reason, sources with emission directly in the telecom C-band (\SI{1530}{\nano\meter} - \SI{1565}{\nano\meter}) become crucial, among other reasons, due to the global loss minimum in standard glass fibers.\\
In this context, recent theoretical \cite{Barbiero2022} and experimental studies \cite{Nawrath2023, Holewa2023} on QDs coupled to circular Bragg grating cavities (CBG) represent a major advance for QD single-photon sources emitting in the telecom C-band. Apart from the single-photon nature, the indistinguishability of emitted photons becomes a key property for interference-based applications. This opens possibilities towards implementations of such as quantum teleportation \cite{Bouwmeester1998,Anderson2020} or the generation of entangled photon states e.g. for enhanced resolution in quantum metrology \cite{Muller2017, Giovannetti2011, Slussarenko2017}. Finally, narrow linewidth of the emitted photons becomes crucial for an interface with some other system, as for instance with a quantum memory \cite{Thomas2023, Thomas2023a}. All of the introduced criteria (triggered single-photons, high brightness and indistinguishability, telecom C-band compatibility and narrow linewidth) are required for the implementation of an efficient single-photon quantum repeater \cite{Sangouard2007}, a pivotal element for long-distance quantum communication.\\
Depending on an actual protocol for a quantum communication application, also the coherence in the photon number basis becomes important \cite{Loredo2019, Bozzio2022, Karli2023}. Some applications, as for instance twin-field QKD rely on coherence in the photon number basis \cite{Lucamarini2018,Karli2023}. On the other hand, the performance of other protocols, e.g. BB84, is affected by this coherence since it opens additional possibilities for attacks \cite{Bozzio2022, Cao2015}.\\\\
In this work, we present a triggered, telecom C-band single-photon source which combines high brightness and indistinguishability with sub-GHz linewidth. In this context, values of $g^{(2)}(0) = \SI{0.026\pm0.001}{}$ ($g^{(2)}(0) = \SI{0.014\pm0.001}{}$), an application-ready rate of \SI{0.75}{\mega\hertz} (\SI{1.45}{\mega\hertz}), a two-photon interference visibility of \SI{0.504\pm0.014}{} (\SI{0.664\pm0.004}{}) and a linewidth of \SI{0.79\pm0.01}{\giga\hertz} (\SI{0.80\pm0.01}{\giga\hertz}) are simultaneously achieved under coherent (incoherent) pumping combined with spectral filtering. To the best of our knowledge, a combination of these characteristics in a single source has been elusive, paving the way towards single-photon quantum repeater implementations.\\
We show the versatility of the source by operation under incoherent and coherent pumping conditions with similar performance which makes it appealing for a broad range of applications. For the coherent pumping conditions we use the novel SUPER (Swing-Up of Quantum EmitteR population) scheme which was only recently proposed \cite{Bracht2021} and realized for QDs emitting in the near-infrared regime \cite{Karli2022, Boos2022}. This work denotes the first implementation of the SUPER scheme with a telecom C-band QD whereas we use an alternative approach for the generation of the synchronized pumping pulses.

\section*{Experimental conditions for (coherent) operation}\label{sec:Setup}
\begin{figure*}
	\centering
	\includegraphics{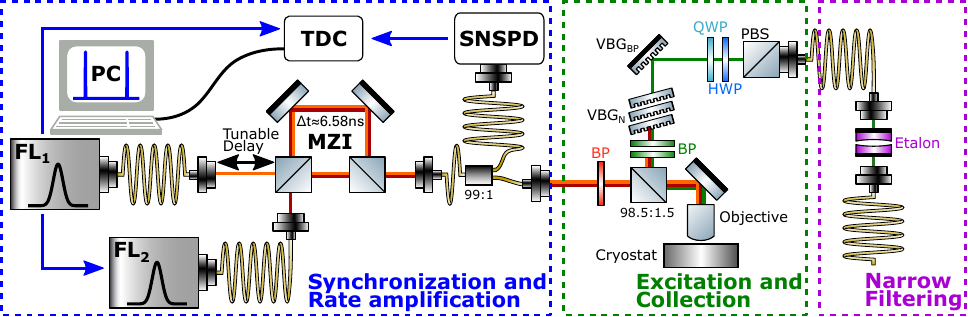}
	\caption{\textbf{Schematic of the setup used for coherent pumping via the SUPER scheme using two synchronized fiber lasers (FLs)}. The pulses of two synchronized FLs are superimposed in an unbalanced Mach-Zehnder interferometer (MZI). Comparing the arrival time of the pulses on a superconducting nanowire single-photon detector (SNSPD) to the synchronization signal of FL$_1$ with a time-to-digital converter (TDC) provides the relative time delay between the pulses which can be adjusted win the tunable delay in the MZI. The emission of the QD-microcavity system pumped by the synchronized pulses is collected with an objective and filtered with a series of volume Bragg grating (VBG) notch (N) and bandpass (BP) filters. The single-photon stream is coupled into a single-mode fiber via fiber coupler attached to a polarizing BS (PBS). Additionally, the signal can be filtered with a narrow-band etalon.}
	\label{fig:1}
\end{figure*}
The source is based on a QD-microcavity system consisting of an In(Ga)As QD emitting in the telecom C-band coupled to a circular Bragg grating cavity. For the measurements, the positively charged exciton state at \SI{1551}{\nano\meter} is employed which exhibits a lifetime of $\sim\SI{500}{\pico\second}$ owing to a Purcell enhancement of \SI{4.3\pm0.8}{} (see supplementary material section I). The emission of the QD-microcavity system is highly polarized ($>\SI{95}{\percent}$) which we ascribe to the small ellipticity of the cavity and an spatial offset of the QD from the cavity center resulting in an uneven coupling to the two cavity polarization modes \cite{Wang2019, Peniakov2023}.\\
Commonly, coherent operation of a QD single-photon source is performed via resonance fluorescence with polarization-based filtering \cite{Muller2007, Benelajla2021}. This comes, however, along with an intrinsic limitation of the achievable efficiency which can only be reduced using specially tailored microcavities \cite{Wang2019, Tomm2021}. In this work, we employ coherent pumping via the recently proposed SUPER scheme \cite{Bracht2021}. There, the QD is pumped with two synchronized laser pulses which are both red-detuned compared to QD transition enabling straight forward spectral filtering. For a given detuning $\Delta_1<0$ of the first laser, maximum population of the excited state can be reached for a detuning the second laser of $|\Delta_2| > 2\cdot|\Delta_1|$ \cite{Bracht2021, Bracht2023}.\\
In previous implementations of the SUPER scheme, it has been shown that the two pulses can be carved out of a single, broadband pulse of a laser with femtosecond (fs) pulse length \cite{Karli2022,Boos2022} resulting in two spectrally detuned picosecond (ps) pulses, as needed for the QD excitation. In this work, we take an alternative route, superimposing independently generated pulses from two separate fiber lasers (FLs) directly operating with pulse lengths in the ps-regime. As the SUPER scheme operates at large detunings in the range of several \SI{}{meV} and with comparably high excitation powers, this approach circumvents possible power limitations due to the intrinsically inefficient pulse-shaping and enables the use of ps-lasers that are commonly employed in typical QD setups. On the other hand, the possibility to synchronize the two laser sources with ps-precision is required.\\
\hyperref[fig:1]{Fig. 1} shows a schematic of the employed setup for that purpose. At first, the two FLs are precisely synchronized based on the arrival times of the pulses on a superconducting nanowire single-photon detector (SNSPD). Alternatively, the synchronization of the FLs can also be achieved by exploiting an intrinsic effect of the excitation scheme when using standard glass fibers: Due to the small non-linearity of optical fibers, a (weak) four-wave mixing (FWM) signal is observed after the synchronization of the FLs. Vice versa, maximizing the FWM signal after the fiber optimizes the pulse overlap due to the strong power dependence of the process (see further information in supplementary material section II).\\
The FWM signal is filtered before reaching the QD whereas the collected single-photon emission is separated from the laser signal with spectral filters (more information in supplementary material section III). Finally, the emission is coupled into a single-mode fiber with an attached polarizing beam splitter (PBS) exploiting the high degree of polarization of the source to generate a polarized photon stream. Employing narrowband filtering with an etalon enables operation conditions of the final source in the sub-GHz regime.

\section*{Coherent pumping}\label{sec:SUPER}
\begin{figure*}
	\centering
	\includegraphics[width = 18cm]{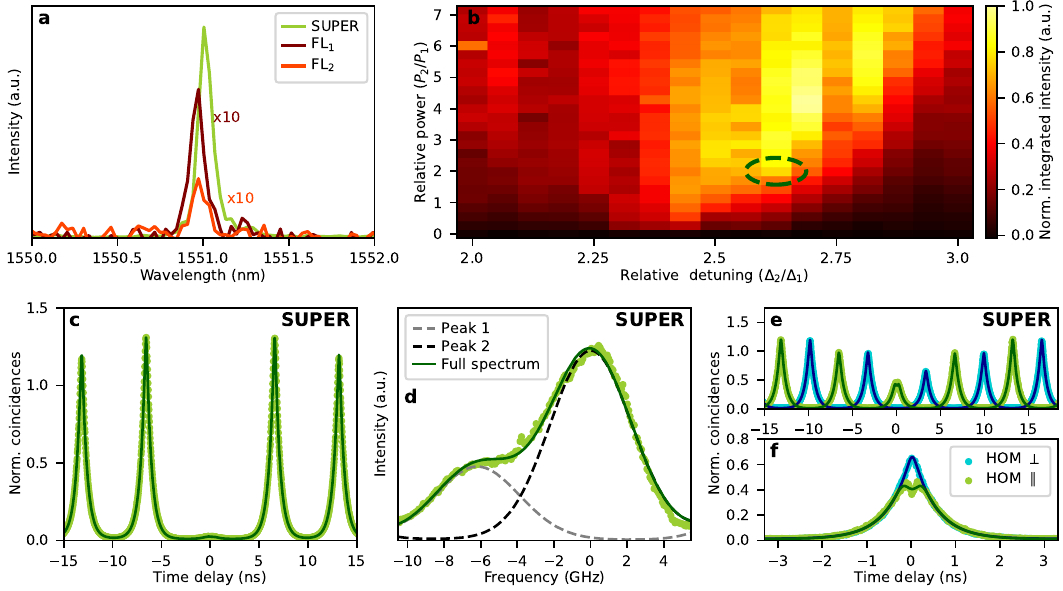}
	\caption{\textbf{Emission characteristics of the QD-microcavity system under coherent SUPER excitation}. \textbf{a}\,Emission spectrum under SUPER pumping and components of the spectrum when excited with only one laser via the phonon sideband. \textbf{b}\,Scan of detuning $\Delta$ and power $P$ of FL$_2$ for $\Delta_1 = \SI{4}{\nano\meter}$ and $P_1 = \SI{350}{\nano\watt}$ (repetition rate $f_\text{rep} = \SI{76}{\mega\hertz}$). Final operation point of the SUPER scheme marked with green ellipse. \textbf{c}\,Measurement of the photon autocorrelation $g^{(2)}(\tau)$ with $g^{(2)}(0) = \SI{0.076\pm0.001}{}$. \textbf{d}\,High-resolution spectrum of the emission line revealing a doublet-peak with single-peak linewidth of $\Gamma_\text{FWHM} = \SI{4.94\pm0.05}{\giga\hertz}$. \textbf{e,f}\,Photon indistinguishability measurement yields a two-photon interference (TPI) visibility $V_\text{TPI} = \SI{0.104\pm0.014}{}$. The orthogonal measurement in panel \textbf{e} is offset horizontally for the sake of visual clarity.}
	\label{fig:2}
\end{figure*}
In order to obtain efficient excitation in the SUPER scheme, it is necessary to investigate the rich parameter space of the two laser pulses. Therefore, we scan the detuning $\Delta$ to the QD transition and power $P$ of FL$_2$ (no time delay between the laser pulses) for a fixed set of parameters of the first laser ($\Delta_1 = \SI{4}{\nano\meter}$, $P_1 = \SI{350}{\nano\watt}$ at a repetition rate of $f_\textbf{rep} = \SI{76}{\mega\hertz}$). An exemplary spectrum is shown in \hyperref[fig:2]{Fig. 2a} representing the emission at the optimal operation conditions (obtained from the scan depicted in \hyperref[fig:2]{Fig 2b}). A single, narrow line at \SI{1551}{\nano\meter} is observed when both lasers are employed, as anticipated for a successful implementation of the SUPER scheme. Interestingly, also with only one of the two lasers active, the transition is still excited, though with strongly reduced intensity. This non-zero emission is ascribed to pumping via the absorption of a phonon (see supplementary material section IV). Nevertheless, both signals are more than an order of magnitude lower than when using the actual SUPER pumping. The result of the scan can be seen in \hyperref[fig:2]{Fig 2b} where the color code corresponds to the integrated intensity of the emission line corrected for the signal component due to the phonon-induced pumping. The resonance around $3\cdot|\Delta_1|>|\Delta_2| > 2.5\cdot|\Delta_1|$ clearly shows the fingerprint of the coherent-pumping in the SUPER scheme \cite{Boos2022}. In order to minimize the influence of the incoherent phonon-induced pumping, we choose the final operation point for the measurements presented in the following at the edge of the resonance at $\Delta_2 = 2.7\cdot\Delta_1$, $P_2 = 2\cdot P_1$ (\SI{10.75}{\nano\meter}, \SI{700}{\nano\watt} at $f_\text{rep} = \SI{76}{\mega\hertz}$). All of the measurement discussed in the following are carried out at $f_\text{rep} = \SI{152}{\mega\hertz}$ (same detuning and power per pulse as for the parameter scan).\\
For any application of a single-photon source, purity and brightness of the emitted photon stream are decisive characteristics. To assess the former, we carry out a measurement of the photon autocorrelation $g^{(2)}(\tau)$ in a Hanbury-Brown and Twiss (HBT) interferometer, as displayed in \hyperref[fig:2]{Fig. 2c}. By comparing the integrated coincidences within one repetition period around zero time delay (only corrected for detector dark counts) to the same quantity of a peak in the Poissonian level yields a good single-photon purity with $g^{(2)}(0) = \SI{0.076\pm0.001}{}$. Based on this value, the detected count rate ($CR_\text{raw} = \SI{6.50}{\mega\hertz}$) and the detection efficiency (\SI{76.8}{\percent}, a detailed description  of the efficiencies can be found in the supplementary material), an application-ready rate of polarized single-photons of $CR_\text{end} = \SI{8.14}{\mega\hertz}$ coupled into a single-mode fiber, corresponding to an end-to-end efficiency of $\eta_\text{end} = \SI{5.36}{\percent}$, is found. Until now, application oriented research on QD-based single-photon sources in telecom C-band has been limited to incoherent excitation schemes while investigations on coherent pumping schemes have put the focus more on fundamental research \cite{Nawrath2019, Anderson2021, Nawrath2021, Wells2022}. However, in terms of application-ready brightness, a rate of \SI{500}{\kilo\hertz} under continuous-wave excitation has not yet been surpassed \cite{Wells2022} underlining the significance of the implementation in this work.\\
Next, the photon indistinguishability is investigated. For QDs under coherent pumping this is mainly affected by spectral diffusion \cite{Nawrath2021,Kuhlmann2015} due to noise in the magnetic and electric field in the QD vicinity \cite{Sallen2010,Kuhlmann2015}. To probe the influence of this broadening on the QD linewidth,
a high-resolution spectrum of the emission line is acquired via a scanning Fabry-P\'erot interferometer, as depicted in \hyperref[fig:2]{Fig. 2d}. Interestingly, the scan reveals a doublet peak with a splitting of \SI{6.19}{\giga\hertz} whereas 'Peak 1' has a relative area of \SI{39}{\percent} of 'Peak 2'. Such a finding can be explained by jumps of the line by the energy of the splitting, e.g. due to a fluctuating occupancy of a defect state in the vicinity of the QD \cite{Houel2012, Hauck2014, Vural2020}. Fitting this spectrum with a sum of two Voigt profiles with a common width yields the introduced splitting and amplitudes as well as a single-peak linewidth of $\Gamma_\text{FWHM} = \SI{4.94\pm0.05}{\giga\hertz}$. As the Gaussian component of the Voigt profile dominates, the Lorentzian component has to be fixed to a value of \SI{0.32}{\giga\hertz} corresponding to the Fourier limit. The influence of the finite resolution of the FPI of \SI{70}{\mega\hertz} is considered via a convolution of the Voigt profile with a Lorentzian with said width. However, the degree of indistinguishability is only affected by the amount of broadening taking place on the time scale of the photon separation \cite{Vural2020}, i.e. \SI{6.6}{\nano\second} for the measurements presented here. This is probed by Hong-Ou-Mandel (HOM) type measurements shown in \hyperref[fig:2]{Fig. 2e,f}. Comparing the central peak areas (integration over all coincidences in one full repetition period) of the measurement with photons with parallel polarization interfering on the BS to the fully distinguishable case (orthogonal polarization) yields a degree of indistinguishability of $V_\text{TPI} = \SI{0.104\pm 0.014}{}$. This value is comparable to previously reported results on similar CBG structures emitting in the telecom C-band \cite{Nawrath2023, Holewa2023}. Integrating the QD layer into a gated structure, thus stabilizing the local electrical field, is expected to decrease the line broadening due to the spectral diffusion \cite{Kuhlmann2015}, consequently improving the photon indistinguishability.

\section*{Incoherent pumping}\label{sec:LA}
\begin{figure*}
	\includegraphics[width = 18cm]{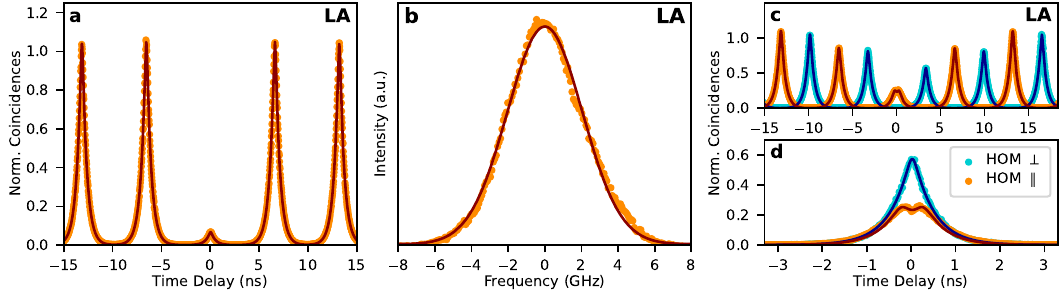}
	\caption{\textbf{Incoherent, phonon-assisted pumping of the QD-microcavity system.} \textbf{a}\,Photon autocorrelation measurement yields $g^{(2)}(0) =  \SI{0.069\pm0.001}{}$. \textbf{b}\,High-resolution spectrum shows a linewidth of $\Gamma_\text{FWHM} = \SI{4.83\pm0.03}{\giga\hertz}$. \textbf{c,d}\,Photon indistinguishability measurement resulting in $V_\text{FPI} = \SI{0.349\pm0.005}{}$. The orthogonal measurement in panel \textbf{c} is offset horizontally for the sake of visual clarity.}
	\label{fig:3}
\end{figure*}
In the following, we operate the QD-microcavity system under incoherent pumping conditions. As we focus in this paper on the high-brightness aspect, primarily spectrally filterable pumping schemes are considered due to the, in general, high intrinsic losses of polarization based filtering \cite{Wang2019}. Here, we choose LA pumping, an incoherent excitation scheme \cite{Vyvlecka2023} with a blue-detuned laser which can combine robustness, high brightness \cite{Glassl2013} and high indistinguishability \cite{Reindl2019}.\\
The measurements are performed at $\Delta = \SI{-2}{\nano\meter}$ and $P = \SI{700}{\nano\watt}$ ($f_\text{rep} = \SI{152}{\mega\hertz}$). The photon autocorrelation measurement under these conditions, depicted in \hyperref[fig:3]{Fig. 3a}, yields a value of $g^{(2)}(0) =  \SI{0.069\pm0.001}{}$ and thus a single-photon purity comparable to SUPER pumping. This is accompanied by brightness measures of $CR_\text{end} = \SI{9.25}{\mega\hertz}$ ($CR_\text{raw} = \SI{7.36}{\mega\hertz}$, $\eta_\text{end} = \SI{6.09}{\percent}$). Thus, similar performance under brightness aspects are achieved under coherent and incoherent pumping. It has to be noted that for LA pumping a reduced blinking behavior is found (QD is optically active in \SI{94}{\percent} (\SI{67}{\percent}) of the time under LA (SUPER)) pumping which contributes to the difference in the brightness measures. The difference is likely related to an altered QD environment for the two pumping schemes which can arise due to the different energies of the pumping lasers.\\
This assumption is further supported by the FPI scan of the emission line in \hyperref[fig:3]{Fig. 3b}. Contrary to the doublet feature under SUPER pumping, only a single peak with a linewidth of $\Gamma_\text{FWHM} = \SI{4.83\pm0.03}{\giga\hertz}$ is observed. This indicates a stable charge configuration of the defect state in the vicinity of the QD preventing the line jumps observed under SUPER excitation. In turn, a more stable charge configuration is expected to reduce blinking due to the suppression of tunneling of charge carriers from QD to defect state \cite{Yang2022}. Again, an integration of the QD layer into a gated structure is expected to further diminish such detrimental effects of local defects \cite{Houel2012}.\\
To finalize the comparison, we perform a HOM measurement under LA excitation, as depicted in \hyperref[fig:3]{Fig. 3c,d} which yields $V_\text{TPI} = \SI{0.349\pm0.005}{}$. This denotes a clear improvement compared to the degree of indistinguishability under SUPER pumping. As the (single-peak) FPI linewidths of the different excitation schemes are similar, this indicates that spectral diffusion (and line jumps) take place on a shorter time scale for the SUPER than for LA. However, this degree of indistinguishability does not only represent an improvement compared to the SUPER scheme but also to previously reported results on InAs/GaAs QDs in the telecom C-band \cite{Nawrath2021, Nawrath2023} and a similar performance as the currently highest reported value across all telecom C-band QD sources \cite{Vajner2023}.

\section*{Narrow-band source with high indistinguishability}\label{sec:Etalon}
\begin{figure*}
	\includegraphics[width = 18cm]{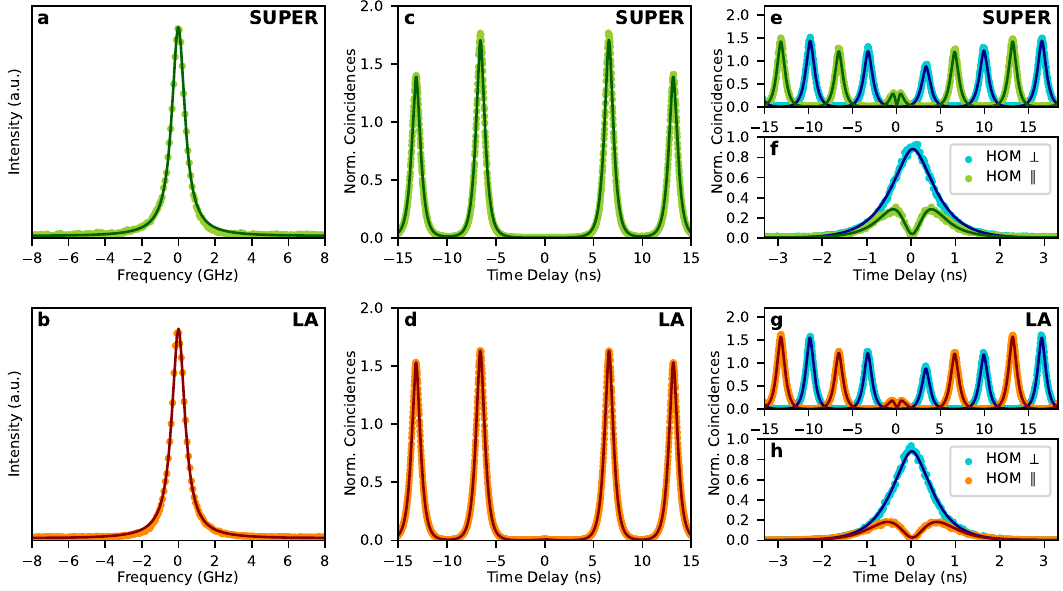}
	\caption{\textbf{Source performance including narrow-band filtering.} SUPER (LA) pumping in the top (bottom) row. \textbf{a,b}\,High-resolution spectra showing Lorentzian lineshapes with $\Gamma_\text{FWHM} = \SI{0.79\pm0.01}{\giga\hertz}$ ($\Gamma_\text{FWHM} = \SI{0.80\pm0.01}{\giga\hertz}$). \textbf{c,d}\,Photon autocorrelation measurements yielding $g^{(2)}(0) = \SI{0.026\pm0.001}{}$ $(g^{(2)}(0) = \SI{0.014\pm0.001}{})$. \textbf{e-h}\,Photon indistinguishability measurements with $V_\text{TPI} = \SI{0.508\pm0.018}{}$ ($V_\text{TPI} = \SI{0.664\pm0.001}{}$). The orthogonal measurements in the panels \textbf{e,g} are offset horizontally for the sake of visual clarity.}
	\label{fig:4}
\end{figure*}
The excellent brightness of the QD-microcavity system allows to employ spectral filtering with a narrow-band etalon in order to further boost the indistinguishability. The lineshape of the source is then governed by the linewidth of the etalon and, consequently exhibits a Lorentzian profile with $\Gamma_\text{FWHM} = \SI{0.79\pm0.01}{\giga\hertz}$ ($\Gamma_\text{FWHM} = \SI{0.80\pm0.01}{\giga\hertz}$) under SUPER (LA) pumping, as depicted in \hyperref[fig:4]{Fig. 4a(b)}. Such a narrow linewidth ($\sim$$2.5$ times the Fourier limit) benefits any interface of the source with other systems, as for instance required for quantum teleportation \cite{Bouwmeester1998,Anderson2020}. This is the case since remote interference does only depend on the measured linewidth and not on the TPI of consecutive photons from one source \cite{Kambs2018}. Especially, narrowband operation of the single-photon source in the sub-GHz regime enables efficient interfacing with atomic memories \cite{Thomas2023, Thomas2023a}.\\
Furthermore, the filtering positively influences the single-photon purity of the source resulting in values of $g^{(2)}(0) = \SI{0.026\pm0.001}{}$ ($g^{(2)}(0) = \SI{0.014\pm0.001}{}$), depicted in \hyperref[fig:4]{Fig. 4c(d)}, under SUPER (LA). Despite the additional losses due to the filtering, the final source exhibits an appealing brightness with  $CR_\text{end} = \SI{0.75}{\mega\hertz}$ ($CR_\text{raw} = \SI{0.58}{\mega\hertz}$, $\eta_\text{end} = \SI{0.49}{\percent}$) under SUPER pumping. As this is partially limited due to the doublet feature in the unfiltered spectrum, even higher rates are observed under LA excitation with $CR_\text{end} = \SI{1.45}{\mega\hertz}$ ($CR_\text{raw} = \SI{1.12}{\mega\hertz}$, $\eta_\text{end} = \SI{0.95}{\percent}$).\\
Finally, the indistinguishability of the source, including the narrow-band filtering, is probed via HOM measurements, as depicted in \hyperref[fig:4]{Fig. 4e-f}. The central peak of the parallel measurements is strongly suppressed for both pumping schemes leading to values of $V_\text{TPI} = \SI{0.508\pm0.018}{}$ ($V_\text{TPI} = \SI{0.664\pm0.004}{}$) for SUPER (LA). To the best of our knowledge, both values represent the highest indistinguishabilities reported so far for QD sources emitting in the telecom C-band. The small discrepancy between the two values is due to different time scale of spectral diffusion mechanisms for the two pumping schemes.

\section*{Conclusion}\label{sec:Conclusion}
In summary, we have shown the first triggered telecom C-band single-photon source that combines high values of purity ($g^{(2)}(0) = \SI{0.026\pm0.001}{}$/$g^{(2)}(0) = \SI{0.014\pm0.001}{}$), brightness (\SI{0.75}{\mega\hertz}/\SI{1.45}{\mega\hertz} application-ready rate) and indistinguishability (two-photon interference visibility of \SI{0.504\pm0.014}{}/\SI{0.664\pm0.004}{}) with narrow linewidth (\SI{0.79\pm0.01}{\giga\hertz}/\SI{0.80\pm0.01}{\giga\hertz}) under coherent/incoherent operation. This constitutes an important step towards prospective implementations of a single-photon quantum repeater where mature sources combining these features are required. The system is based on an In(Ga)As/GaAs QD coupled to a circular Bragg grating cavity combined with efficient narrowband spectral filtering.\\
The source is operated under coherent as well as incoherent pumping conditions with similar performance. This versatility in the operation conditions opens the way for a broad range of applications benefitting from the presence or absence of coherence in the photon number states. Coherent pumping is implemented by the first realization of the SUPER scheme for QDs emitting in the telecom C-band. In this way, emission with state-of-the-art brightness (application-ready rate of \SI{8.14}{\mega\hertz} without narrowband filtering) under coherent pumping is achieved, whereas previous high brightness QD sources in the telecom C-band were only operated under incoherent pumping. 

\section*{Methods}\label{sec:Methods}
For synchronization of the laser pulses in the SUPER scheme, we use two fiber lasers (FL) from PriTel, namely the models FFL (FL$_1$) and UOC (FL$_2$), which are close in their natural repetition rate. This facilitates to use the electrical signal of FL$_1$ as clock for the lasing operation of FL$_2$ leading to synchronized operation of the lasers. In order to set the relative time delay between the pulses from the different lasers, a superconducting nanowire single-photon detector (SNSPD, model Single Quantum Eos) and a time-to-digital-converter (TDC, model Swabian instruments Time Tagger Ultra) are employed. The relative time delay between both lasers is determined by comparing the arrival time of each FLs pulses on the SNSPD to the electrical pulse of FL$_1$ with the TDC. As introduced in the main text, also maximization of a four-wave mixing signal can be used as indicator for the overlap between the pulses. This time delay is adjusted by tuning the optical path length of the pulses from FL$_1$.\\
As both lasers operate at a repetition rate of $\SI{76}{\mega\hertz}$ an unbalanced Mach-Zehnder interferometer is used to double the repetition rate.

\backmatter

\bmhead{Supplementary information}
Additional information to the presented work is online available.

\bmhead{Acknowledgments}
The authors like to thank Julian Maisch and Michal Vyvlecka for fruitful discussions and the solid-state quantum photonics group of the IFW Dresden (in person of Caspar Hopfmann) as well as the institute of solid state physics of the Leibniz Universit\"at Hannover (in person of Fei Ding) for technical support during the measurements.

\bmhead{Funding}
The authors gratefully acknowledge funding by the German Federal Ministry of Education and Research (BMBF) via the project QR.X (No.16KISQ013) and the European Union's Horizon 2020 research and innovation program under Grant Agreement No. 899814 (Qurope). Furthermore, this project (20FUN05 SEQUME) has received funding from the EMPIR programme co-financed by the Participating States and from the European Union's Horizon 2020 research and innovation programme. This work was also funded by the Deutsche Forschungsgemeinschaft (DFG, German Research Foundation) - 431314977/GRK2642 and the Quantum Technology BW via the project TelecomSPS.

\bmhead{Contributions}
R.J. performed the measurements with support of W.F. and C.N.. R.J. analyzed the data with the help of W.F.. S.B., C.R. and S.K. fabricated the circular Bragg grating microcavities. P.V. and R.S. grew the sample with supervision of M.J.. R.J. wrote the manuscript with the help of C.N., S.L.P and P.M.. S.L.P. and P.M. supervised the project. All authors contributed to scientific discussions.

\bmhead{Data availability}
Data supporting the findings presented in this work are available from the authors upon reasonable request.

\bmhead{Competing interests}
The authors declare no competing interests.





\begin{appendices}

\end{appendices}


\end{document}